\documentclass[preprint,aps,amsmath,superscriptaddress,nofootinbib,tightenlines,showpacs]{revtex4}
\usepackage{txfonts}
\usepackage{mathrsfs}
\usepackage{amsmath}
\usepackage{amsfonts}
\usepackage{amssymb}
\usepackage{latexsym}
\usepackage{graphicx}
\usepackage{bm}
\usepackage{epsfig}
\usepackage[english]{babel}
\usepackage{comment}
\usepackage{float}
\def \be {\begin{equation}}
\def \ee {\end{equation}}
\def \bea {\begin{eqnarray}}
\def \eea {\end{eqnarray}}
\def \nn {\nonumber}

\def \a {\alpha}
\def \b {\beta}

\def \d {\delta}

\def \m {\mu}
\def \n {\nu}
\def \k {\kappa}

\def \s {\sigma}
\def \r {\rho}
\def \o {\omega}

\def \th {\theta}
\def \Th {\Theta}

\def \t {\tau}
\def \dag {\dagger}
\def \p {\partial}

\def\bd{\begin{document}}
\def\ed{\end{document}}
\def\nn{\nonumber}
\def\bea{\begin{eqnarray}}
\def\eea{\end{eqnarray}}
\let\bm=\bibitem
\let\la=\label

\def\N{{\cal N}}
\def\sst{\scriptscriptstyle}
\def\thetabar{\bar\theta}
\def\Tr{{\rm Tr}}
\def\one{\mbox{1 \kern-.59em {\rm l}}}

\def\a{\alpha}      \def\da{{\dot\alpha}}
\def\b{\beta}       \def\db{{\dot\beta}}
\def\c{\gamma}  \def\C{\Gamma}  \def\cdt{\dot\gamma}
\def\d{\delta}  \def\D{\Delta}  \def\ddt{\dot\delta}
\def\e{\epsilon}        \def\vare{\varepsilon}
\def\f{\phi}    \def\F{\Phi}    \def\vvf{\f}
\def\h{\eta}
\def\k{\kappa}
\def\l{\lambda} \def\L{\Lambda}
\def\m{\mu} \def\n{\nu}
\def\o{\omega}
\def\P{\Pi}
\def\r{\rho}
\def\s{\sigma}  \def\S{\Sigma}
\def\t{\tau}
\def\th{\theta} \def\Th{\Theta} \def\vth{\vartheta}
\def\X{\Xeta}
\def\z{\zeta}
\def\w{\wedge}
\def\u{\underline}
\def\hs{\hspace}

\def\cA{{\cal A}} \def\cB{{\cal B}} \def\cC{{\cal C}}
\def\cD{{\cal D}} \def\cE{{\cal E}} \def\cF{{\cal F}}
\def\cG{{\cal G}} \def\cH{{\cal H}} \def\cI{{\cal I}}
\def\cJ{{\cal J}} \def\cK{{\cal K}} \def\cL{{\cal L}}
\def\cM{{\cal M}} \def\cN{{\cal N}} \def\cO{{\cal O}}
\def\cP{{\cal P}} \def\cQ{{\cal Q}} \def\cR{{\cal R}}
\def\cS{{\cal S}} \def\cT{{\cal T}} \def\cU{{\cal U}}
\def\cV{{\cal V}} \def\cW{{\cal W}} \def\cX{{\cal X}}
\def\cY{{\cal Y}} \def\cZ{{\cal Z}}

\def\ua{\underline{\alpha}} \def\ubb{\underline{\beta}}
\def\ug{\underline{\gamma}}
\def\ub{\underline{\phantom{\alpha}}\!\!\!\beta}
\def\uc{\underline{\phantom{\alpha}}\!\!\!\gamma}
\def\um{\underline{\mu}} \def\un{\underline{\nu}}
\def\ud{\underline\delta}
\def\ue{\underline\epsilon}
\def\una{\underline a}\def\unA{\underline A}
\def\unb{\underline b}\def\unB{\underline B}
\def\unc{\underline c}\def\unC{\underline C}
\def\und{\underline d}\def\unD{\underline D}
\def\une{\underline e}\def\unE{\underline E}
\def\unf{\underline{\phantom{e}}\!\!\!\! f}\def\unF{\underline F}
\def\unm{\underline m}\def\unM{\underline M}
\def\unn{\underline n}\def\unN{\underline N}
\def\unp{\underline{\phantom{a}}\!\!\! p}\def\unP{\underline P}
\def\unq{\underline{\phantom{a}}\!\!\! q}
\def\unQ{\underline{\phantom{A}}\!\!\!\! Q}
\def\unH{\underline{H}}
\def\ul{\underline}

\def\As {{A \hspace{-6.4pt} \slash}\;}
\def\bs {{b \hspace{-6.4pt} \slash}\;}
\def\Ds {{D \hspace{-6.4pt} \slash}\;}
\def\ds {{\del \hspace{-6.4pt} \slash}\;}
\def\ss {{\s \hspace{-6.4pt} \slash}\;}
\def\ks {{ k \hspace{-6.4pt} \slash}\;}
\def\ps {{p \hspace{-6.4pt} \slash}\;}
\def\pas {{{p_1} \hspace{-6.4pt} \slash}\;}
\def\pbs {{{p_2} \hspace{-6.4pt} \slash}\;}

\def\Fh{\hat{F}}
\def\Vh{\hat{V}}
\def\Xh{\hat{X}}
\def\ah{\hat{a}}
\def\xh{\hat{x}}
\def\yh{\hat{y}}
\def\ph{\hat{p}}
\def\xih{\hat{\xi}}

\def\psit{\tilde{\psi}}
\def\Psit{\tilde{\Psi}}
\def\tht{\tilde{\th}}

\def\At{\tilde{A}}
\def\Qt{\tilde{Q}}
\def\Rt{\tilde{R}}
\def\Nt{\tilde{N}}

\def\at{\tilde{a}}
\def\st{\tilde{s}}
\def\ft{\tilde{f}}
\def\pt{\tilde{p}}
\def\qt{\tilde{q}}
\def\vt{\tilde{v}}
\def\nt{\tilde{n}}

\def\delb{\bar{\partial}}
\def\bz{\bar{z}}
\def\bD{\bar{D}}
\def\bB{\bar{B}}

\def\bk{{\bf k}}
\def\bl{{\bf l}}
\def\bp{{\bf p}}
\def\bq{{\bf q}}
\def\br{{\bf r}}
\def\bx{{\bf x}}
\def\by{{\bf y}}
\def\bR{{\bf R}}
\def\bV{{\bf V}}

\def\d{\delta}\def\D{\Delta}\def\ddt{\dot\delta}

\def\p{\partial} \def\del{\partial}
\def\xx{\times}
\def\uno{\mbox{1 \kern-.59em {\rm l}}}

\def\trp{^{\top}}
\def\inv{^{-1}}
\def\dag{{^{\dagger}}}
\def\pr{\prime}

\def\rar{\rightarrow}
\def\lar{\leftarrow}
\def\lrar{\leftrightarrow}

\def\cw{{\cal W}}
\def\cz{{\cal Z}}
\def\tcm{\tilde{\cal M}}
\def\sgn{{\rm sgn}}
\def\sd {d^{4|4}}
\def\lan{\langle}
\def\ran{\rangle}

\def\tr{\mbox{tr}}
\def\sign{\mbox{sign}}
\def\fnl{f_\text{NL}}
\def\horava{Ho\v{r}ava}
\def\la{\langle}
\def\ra{\rangle}
\def\mb{\mathbf}
\def\nn{\nonumber}
\def\hl{Ho\v{r}ava-Lifshitz}
\def\p{\partial}
\def\dij{\delta_{ij}}
\def\tr{\mbox{tr}}
\def\sign{\mbox{sign}}
\def\fnl{f_\text{NL}}
\def\horava{Ho\v{r}ava}
\def\la{\langle}
\def\ra{\rangle}
\def\mb{\mathbf}
\def\nn{\nonumber}
\def\hl{Ho\v{r}ava-Lifshitz}
\def\p{\partial}
\def\dij{\delta_{ij}}

\begin{document}

\title{The Cosmological Constant as a Function of Extrinsic Curvature and Spatial Curvature}
\author{Jin-Zhang Tang\footnote{Electronic address:JinzhangTang@pku.edu.cn},\hspace{2ex} Qiang Xu\footnote{Electronic address:xuqiang@pku.edu.cn}} \affiliation{Department of Physics,
and State Key Laboratory of Nuclear Physics and Technology, Peking
University, Beijing 100871, China}

\date{\today\\ \vspace{1cm}}

\begin{abstract}
In this paper we suppose that the cosmological constant will change when the universe expends. For a general consideration, the cosmological constant is assumed to be a function of scale factor and Hubble constant. According to the ADM formulation, to the FRW metric, the extrinsic curvature $I$ equals $-6H^{2}$ and spatial curvature $R$ equals
$6k/a^{2}$. Therefore we suppose cosmological constant is a function of extrinsic curvature and spatial curvature. We investigate the cosmological evolution of FRW universe in these models. At last we investigate two particular models which could fit the observation data about dark energy well. Actually a changeless cosmological constant is not essential. If when the universe expands, the cosmological constant changes slowly and gradually flows to a constant, the observation data about dark energy could also be fitted well by this kind of theory.

\pacs{98.80.Cq}
\end{abstract}
\maketitle

\section{introduction}\label{sec-intro}

Eintein's general relativity(GR) has been considered as a
fundamental theory of gravity. Even from an effective field theory
point of view, it should describe all large scale gravitational
physics, in particular the evolution of our universe. However, the
discovery of dark matter and dark energy from various observation
posed great challenge to the theory. To address the dark energy
issues, the most simplest method is adding a cosmological constant which is called
$\Lambda$CDM model. The cosmological constant problem has been investigated by many papers such as \cite{CC-Weinberg,CC-carroll,CC-Peebles} and so on. Many other theories also have been proposed for
 the cosmic acceleration, for instance, Quintessence\cite{quintessence-01,quintessence-02,quintessence-03}, K-essence\cite{K-essence-01,K-essence-02}, Phantom\cite{Phantom01,Phantom02}, etc.
And numerous versions of modification or extension of GR have
been proposed in the last and this century. The $f(R^{(4)})$
theory \cite{fR01,fR02,fR03,fR04} is one of the
modifications to GR. In this theory, the Lagrangian density $f$
is an arbitrary function of $R^{(4)}$. The theory can explain the
cosmic acceleration without introducing a cosmological constant,
even though the theory would be very much restricted by the solar
system test.

In this paper, we suppose the cosmological constant will change when the universe expends. For the
scalar factor and Hubble constant are main variable to characterize a universe, the cosmological constant
is supposed to be a function of them.
In the ADM formulation \cite{ADM}, the metric is
\begin{equation}
ds^{2}_{4}=-N^{2}dt^{2}+g_{ij}(dx^{i}-N^{i}dt)(dx^{i}-N^{i}dt).
\end{equation}
the building block to construct the action is
the gauge invariant quantity: the extrinsic curvature tensor
\begin{equation}
K_{ij}=\frac{1}{2N}\left(\dot{g}_{ij}-\nabla_{i}N_{j}-\nabla_{j}N_{i}\right),
\end{equation}
where a dot denotes a derivative with respect $t$, and the
3-dimensional spatial curvature $R$ comes from $g_{ij}$. In terms of them,
the action of GR with a cosmological constant is
\begin{equation}
S_{EH}=\frac{1}{16\pi G}\int
dtd^{3}x\sqrt{g}N\left[K_{ij}K^{ij}-K^{2}+R-2\Lambda\right],
\end{equation}
where $K=g^{ij}K_{ij}$. And we define extrinsic curvature $I$ as
\begin{equation}
I\equiv K_{ij}K^{ij}-K^{2}.
\end{equation}

The homogeneous and isotropic universe is described by the FRW metric,
\begin{equation}
ds^{2}=-dt^{2}+a(t)^{2}\left\{\frac{dr^{2}}{1-kr^{2}}+r^{2}\left(d\theta^{2}+\sin^{2}{\theta}d\phi^{2}\right)\right\},
\end{equation}
where $k=0,\pm1$. To this FRW metric, After some trivial calculations, it is easy to get the extrinsic curvature $I=-6H^{2}$ and spatial curvature $R=6k/a^{2}$. Obviously $I$ is proportional to the  square of Hubble constant and $R$ is proportional to the inverse square on scalar factor $a$. Therefore we suppose the cosmological constant $\Lambda=\Lambda(I,R)$ as a function of extrinsic curvature $I$ and spatial curvature $R$.

For the calculational convenience, we define $F(I,R)\equiv I+R-2\Lambda(I,R)$ and the action of this theory is
\begin{equation}\label{action-oringin}
S_{F}=\frac{1}{16\pi G}\int
dtd^{3}x\sqrt{g}NF\left(I,R\right)
\end{equation}
This theory could be regard as an extension of $f(R^{(4)})$ gravity model. In the next section, we exhibit the basic equations by varying the action with respect to the function $N,N_{k},g_{ij}$. Then we investigate the cosmological evolution of FRW universe in these models. At last, we investigate two particular models which could fit the observation data about dark energy well. Actually a changeless cosmological constant is not essential. If the cosmological constant changes slowly and gradually flows to a constant, the observation data about dark energy could also be fitted well by this kind of theory.

\section{Basic Equations}

Let us start from the action
\begin{equation}\label{action-oringin01}
S_{F}=\frac{1}{16\pi G}\int
dtd^{3}x\sqrt{g}NF\left(I,R\right).
\end{equation}
 Actually the action
could be more general if $F$ is an arbitrary function of ``$\hspace{2ex}g_{ij}, K, K_{ij},\hspace{4ex}$ $\nabla_{i}K_{jk},\cdots,\nabla_{i_{1}}\nabla_{i_{2}}\cdots\nabla_{i_{n}}K_{jk}
,\cdots,R,R_{ij},R_{ijkl},\nabla_{i}R_{jk},\cdots$''. In this paper we would like to focus on the simplest case
$F=F\left(I,R\right)$.
Now we consider the case that there are matters couple to the gravity.
The action should be
\begin{equation}\label{action-oringin-matter}
S=S_{F}+S_{m}=\int dtd^{3}x\sqrt{g}N\left[\frac{1}{16\pi
G}F\left(I,R\right)+\mathcal{L}_{m}\right].
\end{equation}
By varying the action with respect to the function
$N,N_{k},g_{ij}$, we get the Hamiltonian constraint,
super-momentum constraint and dynamical equations,
\begin{eqnarray}
\label{Hamiltonian-constraint}
J_{N}&=&\frac{1}{16\pi G}\left[F-2F_{,I}I\right],\\
\label{super-momentum-constraint}
J^{k}&=&\frac{1}{8\pi G}\nabla_{j}\left[F_{,I}\pi^{jk}\right],\\
\label{dynamical-equations} \nn
-\frac{1}{2}p^{ij}&=&\frac{1}{16\pi G}\left\{-\frac{1}{\sqrt{g}N}\partial_{t}\left[\sqrt{g}F_{,I}\pi^{ij}\right]\right.\\
\nn
&+&\left.\frac{1}{N}\nabla_{k}\left[F_{,I}\left(\pi^{ij}N^{k}-\pi^{kj}N^{i}-\pi^{ik}N^{j}\right)\right]-2F_{,I}\left[K^{i}_{m}K^{mj}-\lambda KK^{ij}\right]\right.\\
&-&\left.F_{,R}R^{ij}+\left(\nabla^{i}\nabla^{j}-g^{ij}\nabla^{2}\right)\left(NF_{,R}\right)+\frac{1}{2}Fg^{ij}\right\}.
\end{eqnarray}
$\pi^{ij}$ in above equations is defined as $\pi^{ij}=K^{ij}-\lambda Kg^{ij}$.
And  $J_{N},J^{k},p^{ij}$ are defined as
\begin{equation}
J_{N}\equiv -\frac{\delta(N\mathcal{L}_{m})}{\delta
N};\hspace{2ex}J^{k}\equiv-N\frac{\delta\mathcal{L}_{m}}{\delta
N_{k}};\hspace{2ex}p^{ij}\equiv\frac{2}{\sqrt{g}}\frac{\delta\left(\sqrt{g}\mathcal{L}_{m}\right)}{\delta
g_{ij}}.
\end{equation}
If the matter could be taken to be the perfect liquid, then
 \be
\rho_{m}=J_{N},J^{k}=0,p^{ij}=p_{m}g^{ij}. \ee The  quantities
$\left(J_{N},J^{i},p^{ij}\right)$ should satisfy the conservation
laws,
\begin{eqnarray}
&&\partial_{t}g_{jk}p^{jk}+\frac{2}{\sqrt{g}}\partial_{t}\left(\sqrt{g}J_{N}\right)+\frac{2N_{k}}{\sqrt{g}N}\partial_{t}\left(\sqrt{g}J^{k}\right)=0,\\
&&\nabla_{k}p^{ik}-\frac{1}{\sqrt{g}N}\partial_{t}\left(\sqrt{g}J^{i}\right)-\frac{N^{i}}{N}\nabla_{k}J^{k}-\frac{J^{k}}{N}\left(\nabla_{k}N^{i}-\nabla^{i}N_{k}\right)=0.
\end{eqnarray}

\section{Cosmological Models}

The homogeneous and isotropic universe is described by the FRW
metric,
\begin{equation}
ds^{2}=-c^{2}dt^{2}+a(t)^{2}\left\{\frac{dr^{2}}{1-kr^{2}}+r^{2}\left(d\theta^{2}+\sin^{2}{\theta}d\phi^{2}\right)\right\},
\end{equation}
where $k=0,\pm1$. The extrinsic curvature of the spatial slices is
\begin{equation}
K_{ij}=\frac{1}{c}\frac{\dot{a}(t)}{a(t)}g_{ij};\hspace{4ex}K=\frac{1}{c}\frac{3\dot{a}(t)}{a(t)}.
\end{equation}
Then $I=-6H^{2}$ where $H\equiv\dot{a}/a$ which is defined as Hubble constant. The spatial slices are of constant-curvature
\begin{equation}
R_{ij}=\frac{2k}{a^{2}(t)}g_{ij};\hspace{4ex}R=\frac{6k}{a^{2}(t)}.
\end{equation}
It can be shown that the momentum constraint
\eqref{super-momentum-constraint} is satisfied identically when
$J^{k}=0$ for the perfect fluid. From the Hamiltonian constraint
\eqref{Hamiltonian-constraint} and the dynamical equation
\eqref{dynamical-equations}, and ``$c$'' is rescaled to unity, we get two cosmological equations as
\begin{eqnarray}
\label{Cos-equation-01}
&&F+12F_{,I}\left(\frac{\dot{a}}{a}\right)^{2}=
16\pi G\rho_{m},\\
\label{Cos-equation-02}
&&2\left[F_{,I}\frac{\ddot{a}}{a}+\dot{F}_{,I}\frac{\dot{a}}{a}\right]
+4F_{,I}\left(\frac{\dot{a}}{a}\right)^{2}
-F_{,R}\frac{2k}{a^{2}}+\frac{1}{2}F=-8\pi Gp_{m},
\end{eqnarray}
with the energy conservation equation
\begin{equation}\label{conservation-equ}
\dot{\rho}_{m}+3\frac{\dot{a}}{a}\left(\rho_{m}+p_{m}\right)=0.
\end{equation}
The FRW equations with the matter and the dark energy could be
rewritten as
\begin{eqnarray}
\label{FRW-01}
&&\left(\frac{\dot{a}}{a}\right)^{2}+\frac{k}{a^{2}}=\frac{8\pi
G}{3}\left(\rho_{m}+\rho_{DE}\right),\\
\label{FRW-02}
&&2\frac{\ddot{a}}{a}+\left(\frac{\dot{a}}{a}\right)^{2}+\frac{k}{a^{2}}=-8\pi
G\left(p_{m}+p_{DE}\right).
\end{eqnarray}
Here we have regarded the nonlinear terms in the equations
\eqref{Cos-equation-01},\eqref{Cos-equation-02} as the dark energy
effectively. Comparing \eqref{FRW-01} to \eqref{Cos-equation-01},
\eqref{FRW-02} to \eqref{Cos-equation-02}, it is easy to get
\begin{eqnarray}
\label{DarkEnergy-rho} &&8\pi
G\rho_{DE}=3\left(1-2F_{,I}\right)\left(\frac{\dot{a}}{a}\right)^{2}+\frac{3k}{a^{2}}-\frac{1}{2}F,\\
\label{DarkEnergy-p} &&8\pi
Gp_{DE}=2\left(F_{,I}-1\right)\frac{\ddot{a}}{a}+2\dot{F}_{,I}\frac{\dot{a}}{a}
+\left(4F_{,I}-1\right)\left(\frac{\dot{a}}{a}\right)^{2}-\left(F_{,R}+\frac{1}{2}\right)\frac{2k}{a^{2}}+\frac{1}{2}F.
\end{eqnarray}
So the equation of state $\omega_{DE}$ is given by
\begin{equation}\label{eqn-omega-DE}
\omega_{DE}=-1+\frac{2\left(F_{,I}-1\right)\dot{H}+2\dot{F}_{,I}H-\left(F_{,R}-1\right)\left(2k/a^{2}\right)}
{3\left(1-2F_{,I}\right)H^{2}+\left(3k/a^{2}\right)-\left(1/2\right)F}.
\end{equation}
From the data of $WMAP+BAO+SN^{b}$, the constraints on $\omega_{DE}$ are $\omega_{DE0}=-0.999^{+0.057}_{-0.056}$ \cite{observation01}. We now define a deceleration parameter
\begin{equation}
q=-\frac{\ddot{a}a}{\dot{a}^{2}}.
\end{equation}
From equations \eqref{FRW-01}and\eqref{FRW-02}, we get
\begin{equation}
q_{0}=-\frac{\rho_{DE0}-(1/2)\rho_{m0}}{\rho_{m0}+\rho_{DE0}-3k/(8\pi Ga_{0}^{2})}=-\Omega_{DE0}+\frac{\Omega_{m0}}{2}\approx-0.592,
\end{equation}
here we have used ``$\Omega_{DE0}=0.728,\Omega_{m0}=0.272$'' \cite{observation01} from the data of $WMAP+BAO+H_{0}$. The critical density of the universe if defined as $\rho_{c}\equiv3H^{2}/8\pi G$. From \eqref{FRW-01} it is easy to get
\begin{equation}
\rho_{c0}=\frac{3H_{0}^{2}}{8\pi G}=\rho_{m0}+\rho_{DE0}-\frac{3k}{8\pi Ga_{0}^{2}}.
\end{equation}
From above, The curvature density is defined as $\rho_{k}\equiv-3k/(8\pi Ga^{2})$. Now we could define
\begin{equation}
\Omega_{k}=\frac{\rho_{k}}{\rho_{c}}=\frac{-k}{a^{2}H^{2}}.
\end{equation}
The data of $WMAP+BAO+SN^{b}$ give the constraints on curvature density, $\Omega_{k0}=-0.0057^{+0.0067}_{-0.0068}$ \cite{observation01}. Obviously, It is probable that $k>0$ and the universe is close.
If the universe is accelerating as power-law that $a(t)\propto t^{y}$ today, it is easy
to get $q_{0}=y(1-y)/y^{2}=-0.592$ and $y=2.451$.

We consider the simple case that
\begin{equation}
-2\Lambda(I,R)=g(I)+f(R).
\end{equation}
From \eqref{DarkEnergy-rho} and \eqref{DarkEnergy-p}, it is easy to get
\begin{eqnarray}\label{rhoDEg(I)}
&&8\pi G\rho_{DE}=g_{,I}I-\frac{1}{2}g(I)-\frac{1}{2}f(R),\\
\label{pDEg(I)}
&&8\pi
Gp_{DE}=2\left(g_{,I}+2g_{,II}I\right)\dot{H}-g_{,I}I-f_{,R}\frac{2k}{a^{2}}+\frac{1}{2}g(I)+\frac{1}{2}f(R).
\end{eqnarray}
The equation of state is
\begin{equation}\label{omegaDEg(I)}
\omega_{DE}=-1+\frac{2\left(g_{,I}+2g_{,II}I\right)\dot{H}-f_{,R}(2k/a^{2})}{g_{,I}I-(1/2)g(I)-(1/2)f(R)}.
\end{equation}

Now we investigate a concrete model $-2\Lambda(I,R)=\alpha e^{\beta R}$, here $R=6k/a^{2}$ and we assume $k>0$. The state equation of this model is
\begin{equation}
\omega_{DE}=-1+\frac{2}{3}\beta R.
\end{equation}
It is easy to get $-0.083<\beta R_{0}<0.087$ from $\omega_{DE0}=-0.999^{+0.057}_{-0.056}$ \cite{observation01}. The coefficient $\alpha$ can be fixed from the equation $8\pi G\rho_{DE0}=(-1/2)\alpha e^{\beta R_{0}}$. From \eqref{conservation-equ} and \eqref{FRW-01}, the evolution equation of $a$ is
\begin{eqnarray}
\nn
\left(\frac{H}{H_{0}}\right)^{2}&=&\Omega_{k0}\left(\frac{a_{0}}{a}\right)^{2}
+\Omega_{DE0}e^{\beta R_{0}\{(a_{0}/a)^{2}-1\}}
+\Omega_{m0}\left(\frac{a_{0}}{a}\right)^{3}+\left(1-\Omega_{DE0}-\Omega_{m0}-\Omega_{k0}\right)\left(\frac{a_{0}}{a}\right)^{4},
\end{eqnarray}
where ``$\Omega_{DE0}=0.73, \Omega_{m0}=0.27, \Omega_{r0}=0.014$ ''\cite{observation01}.
The details about the universe evolution are shown in Fig. 1 and 2. Obviously in Fig. 1, the evolutions of deceleration parameter $q$ for $\beta R_{0}= -0.01$ or $0.01$ are nearly identical to the case $\beta R_{0}=0$ which
corresponds to the $\Lambda$CDM model. Fig. 2 shows that the state function $\omega_{DE}$ flow to $-1$ quickly for the cases $\beta R_{0}=-0.01$ or $\beta R_{0}=0.01$ when $\omega_{DE}$ identically equal to $-1$ for $\beta R_{0}=0$.

 \begin{figure}[H]
 \centering
 \includegraphics[width=0.75\textwidth,height=0.30\textheight]{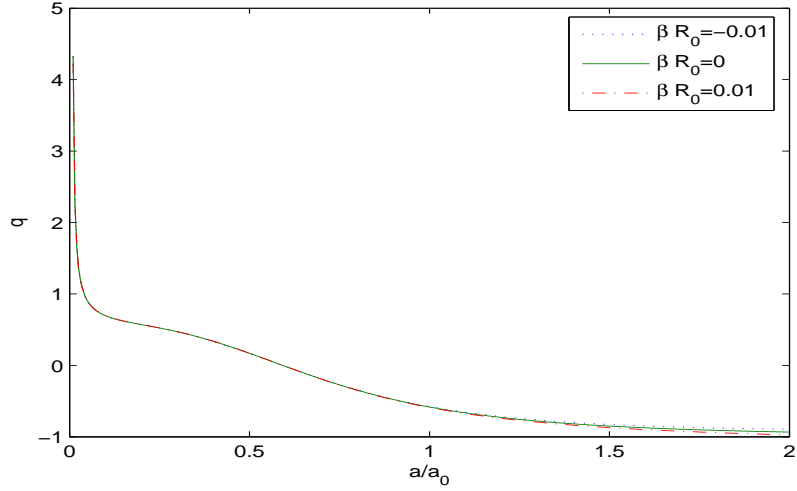}
 \caption{The deceleration parameter $q=-\ddot{a}a/\dot{a}^{2}$ as a function of $a$.}
 \label{fig1}
 \end{figure}

 \begin{figure}[H]
 \centering
 \includegraphics[width=0.75\textwidth,height=0.30\textheight]{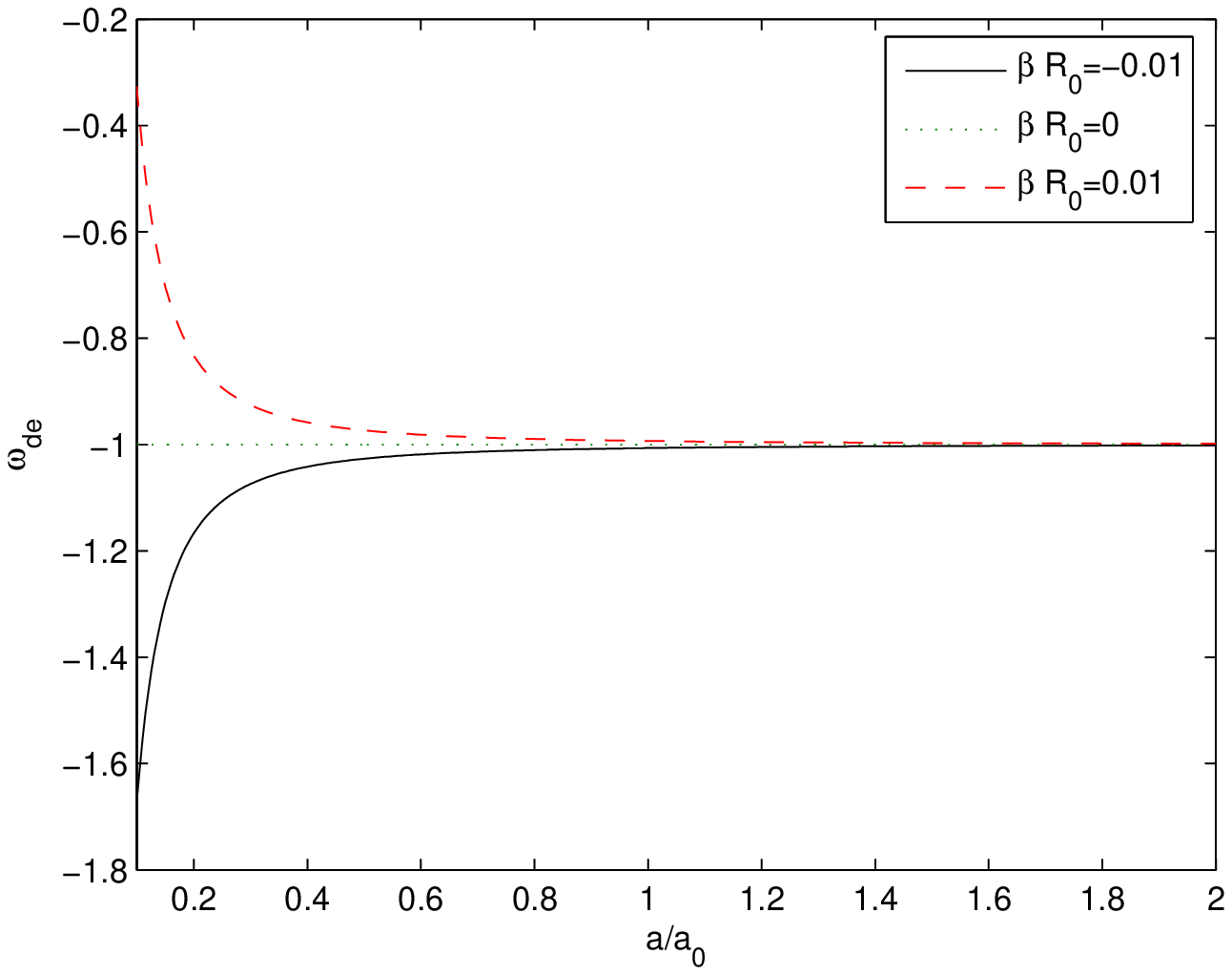}
 \caption{The parameter $\omega_{DE}$ as a function of $a$.}
 \label{fig2}
 \end{figure}

Another specific model is $-2\Lambda(I,R)=\gamma e^{\eta I}$. The corresponding state equation is
\begin{equation}
\omega_{DE}=-1+\frac{2\eta I}{3}\frac{1+2\eta I}{1-2\eta I}\frac{\dot{H}}{H^{2}}.
\end{equation}
When $\omega_{DE0}=-1$, $\eta$ has two solutions $\eta=0$ and $\eta I_{0}=-1/2$. When $\eta=0$, it just is a $\Lambda CDM$ model. The attention will be put on the second case $\eta I_{0}=-1/2$. The coefficient $\gamma$ could be fixed from
\begin{equation}
8\pi G\rho_{DE0}=\gamma\eta e^{\eta I_{0}}I_{0}-\frac{1}{2}\gamma e^{\eta I_{0}},
\end{equation}
and $\gamma=-3e^{1/2}\Omega_{DE0}H_{0}^{2}$. The evolution equation of $a$ is
\begin{eqnarray}
\nn
\left(\frac{H}{H_{0}}\right)^{2}&=&\Omega_{k0}\left(\frac{a_{0}}{a}\right)^{2}
+\frac{1}{2}\Omega_{DE0}\left[1+\left(\frac{H}{H_{0}}\right)^{2}\right]e^{\frac{1}{2}[1-(H/H_{0})^{2}]}
+\Omega_{m0}\left(\frac{a_{0}}{a}\right)^{3}+\left(1-\Omega_{DE0}-\Omega_{m0}-\Omega_{k0}\right)\left(\frac{a_{0}}{a}\right)^{4}.
\end{eqnarray}
Similar to the discussion above, Fig. 3 is the evolution of deceleration parameter $q$ which is similar to $\Lambda$CDM model. Fig. 4 shows that the state function $\omega_{DE}$ flow to $-1$ quickly.

 \begin{figure}[H]
 \centering
 \includegraphics[width=0.75\textwidth,height=0.30\textheight]{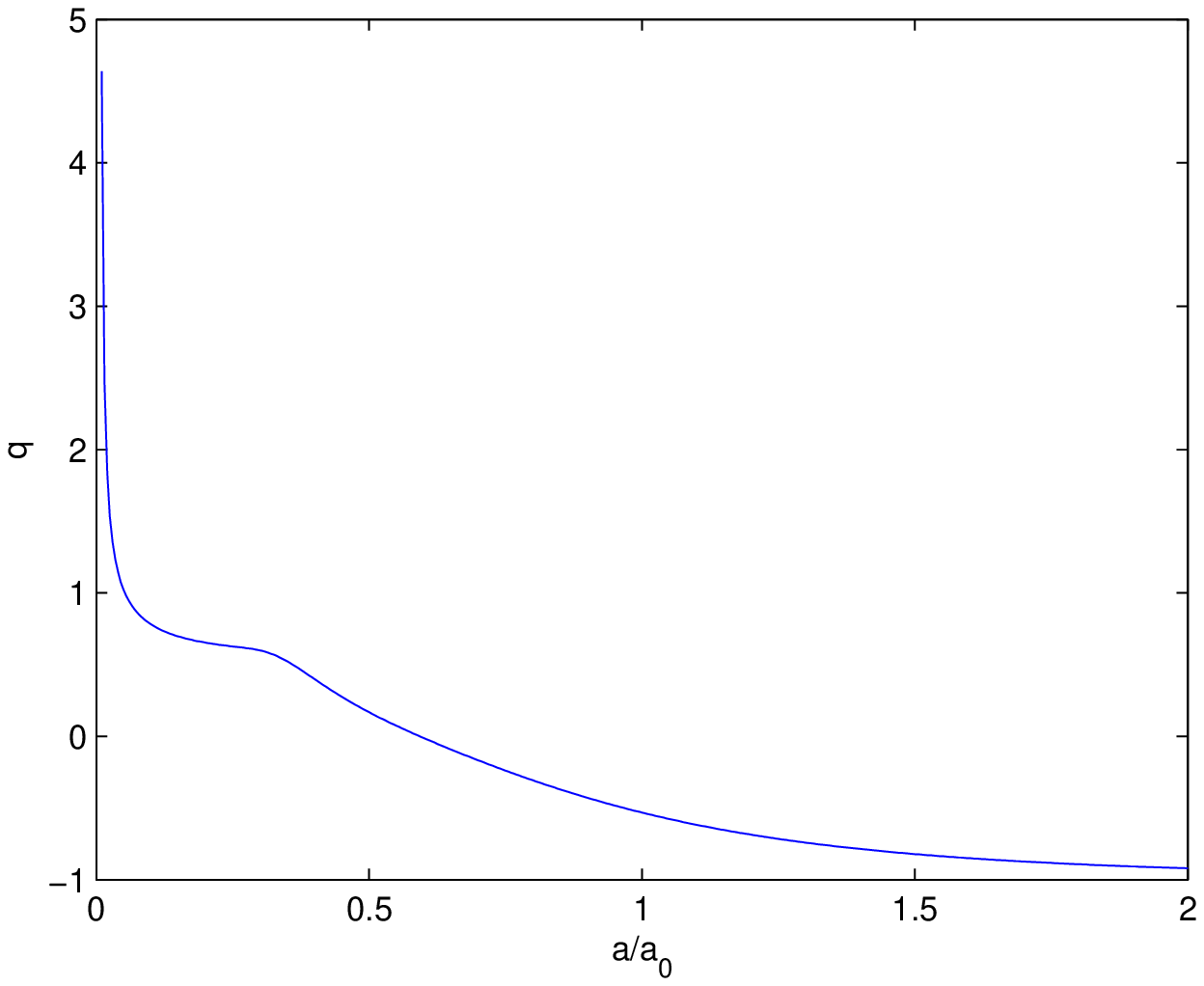}
 \caption{The deceleration parameter $q=-\ddot{a}a/\dot{a}^{2}$ as a function of $a$.}
 \label{fig3}
 \end{figure}

 \begin{figure}[H]
 \centering
 \includegraphics[width=0.75\textwidth,height=0.30\textheight]{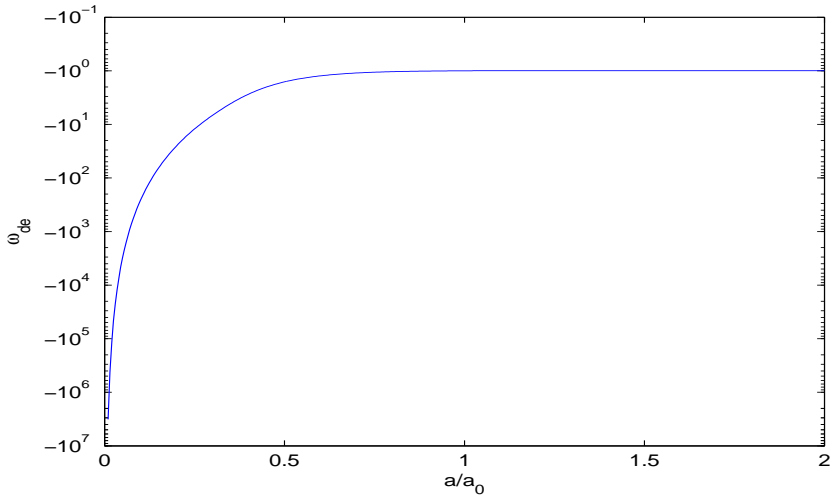}
 \caption{The parameter $\omega_{DE}$ as a function of $a$.}
 \label{fig4}
 \end{figure}

The two models discussed above could be regard as the modification to the $\Lambda$CDM model. To address the dark energy issues, a changeless cosmological constant is not essential. If the cosmological constant changes slowly and gradually flows to a constant, the observation data about dark energy could also be fitted well by this kind of theory.

\section*{Acknowledgments}
The work was partially supported by NSFC Grant No. 10775002,
10975005 and RFDP. We would like to thank Professor Bin Chen very much for
some useful suggestions and his works on the modification of this paper.


\begin{thebibliography}{99}

\bibitem{ADM}
R.L. Arnowitt, S. Deser and C.W. Misner,\emph{The dynamics of
general relativity},``Gravitation:an introduction to current
research'', Louis Witten ed.(Wilew 1962),chapter 7,pp 227-265,
[arXiv:gr-qc/0405109].

\bibitem{CC-Weinberg}
S. Weinberg, ``The cosmological constant problem'', Rev.\ Mod.\ Phys. {\bf 61}, 1(1989).

\bibitem{CC-carroll}
Sean M. Carroll, ``The Cosmological Constant'',  LivingRev.\ Rel.\ 4:1,2001, [arXiv:astro-ph/0004075v2].

\bibitem{CC-Peebles}
 P. J. E. Peebles, Bharat Ratra, ``The Cosmological Constant and Dark Energy '', Rev.\ Mod.\ Phys. {\bf 75: 559-606}(2003), [arXiv:astro-ph/0207347v2].

\bibitem{quintessence-01}
 Ivaylo Zlatev, Limin Wang, Paul J. Steinhardt, ``Quintessence, Cosmic Coincidence, and the Cosmological Constant
'', Phys.\ Rev.\ Lett. {\bf 82:896-899}(1999),[arXiv:astro-ph/9807002v2].

\bibitem{quintessence-02}
A. Yu. Kamenshchik, U. Moschella, V. Pasquier, `` An alternative to quintessence'', Phys.\ Lett.\ B {\bf 511:265-268} (2001), [arXiv:gr-qc/0103004v2].

\bibitem{quintessence-03}
Sean M. Carroll, ``Quintessence and the Rest of the World'', Phys.\ Rev.\ Lett. {\bf 81: 3067-3070}(1998), [arXiv:astro-ph/9806099v2].

\bibitem{K-essence-01}
C. Armendariz-Picon, V. Mukhanov, Paul J. Steinhardt, ``Essentials of k-essence'', Phys.\ Rev.\ D {\bf 63:103510}(2001), [arXiv:astro-ph/0006373v1].

\bibitem{K-essence-02}
Takeshi Chiba, ``Tracking K-essence'', Phys.\ Rev.\ D {\bf 66}(2002), [arXiv:astro-ph/0206298v2].

\bibitem{Phantom01}
 R.R. Caldwell, Phys.Lett.B {\bf 545:23-29}(2002), [arXiv:astro-ph/9908168v2].

\bibitem{Phantom02}
Robert R. Caldwell, Marc Kamionkowski, Nevin N. Weinberg, ``Phantom Energy and Cosmic Doomsday'', Phys. Rev. Lett. {\bf 91}(2003) 071301, [arXiv:astro-ph/0302506v1].



\bibitem{fR01}
S. Nojiri and S. D. Odintsov, eConf {\bf C0602061}, 06 (2006)[Int.
J. Geom. Meth. Mod. Phys. {\bf 4}, 115
(2007)][arXiv:hep-th/0601213].

\bibitem{fR02} T. P. Sotiriou and V. Faraoni, arXiv:0805.1726
[gr-qc].

\bibitem{fR03} A. de Felice and S. Tsujikawa,
arXiv:1002.4928[hep-th].

\bibitem{fR04} Sean M. Carroll, Vikram Duvvuri, Mark Trodden, Michael S.
Turner, ``Is Cosmic Speed-Up Due to New Gravitational Physics?'',
Phys.\ Rev.\  D {\bf 70}, 043528 (2004)[arXiv:astro-ph/0306438].

\bibitem{observation01}  E. Komatsu \emph{et al.}, ``Seven-Year Wilkinson Microwave Anisotropy Probe(WMAP) Observations: Cosmological Interpretation'' , [arXiv:astro-ph/1001.4538v2].















\end{thebibliography}
\end{document}